\documentclass[aps,prb,twocolumn]{revtex4-2}
\usepackage{color}
\usepackage{graphicx}
\usepackage{cancel}
\usepackage{amsfonts}
\usepackage{amssymb}
\usepackage{amsthm}
\usepackage{appendix}
\usepackage{ulem}
\usepackage{url}
\usepackage{hyperref}
\usepackage{float}
\usepackage{amsmath}
\usepackage{multirow}  
\usepackage{dcolumn}   
\usepackage{bm}        
\usepackage{colortbl}
\usepackage{enumerate}
\usepackage{mathtools}
\usepackage{color}
\usepackage{tabularray}

\usepackage[dvipsnames]{xcolor}
\hypersetup{
	colorlinks,
	linkcolor={green!50!black},
	urlcolor={blue!80!black},
	citecolor   = {red!50!black} 
}

\def    \v          {\mbox{$\widetilde{v}$}}

\begin{document}

\title{Effect of Fe-doping on $VS_{2}$ monolayer: A first-principles study}

\author{Mirali Jafari$^{1*}$, Nasim Rahmani-Ivriq$^{2}$, Anna Dyrdał$^{1}$}
\email[e-mail:]{mirali.jafari@amu.edu.pl}
\email[e-mail:]{adyrdal@amu.edu.pl}
\affiliation{$^{1}$Faculty of Physics and Astronomy, ISQI, Adam Mickiewicz University in Poznań, ul. Uniwersytetu Poznanskiego 2, 61-614 Poznan, Poland\\
$^{2}$Materials Simulation Laboratory, Department of Physics, Iran University of Science and Technology, Tehran 1684613114, Iran}

\begin{abstract}
 Transition metal dichalcogenides (TMDs), like $VS_2$, display unique electronic, magnetic, and optical properties, making them promising for spintronic and optoelectronic applications.  Using first-principles calculations based on the Density Functional Theory (DFT), we study the effect of Fe-doping on the electronic and magnetic properties of a $VS_2$ monolayer. The pristine $VS_2$ monolayer has ferromagnetic order and a small energy bandgap. This work aims to comprehensively study the substitution of selected  Vanadium atoms in the $VS_2$ monolayer by Iron (Fe) atoms,  where the substitution concerns  Vanadium atoms at various sites within the 2x2 and 3x3 supercells. This leads to significant modifications of the electronic band structure, magnetic anisotropy energy (MAE), and optical response (e.g., dielectric constant and absorption coefficient). The results provide valuable insights into engineering the $VS_2$ monolayer properties for future applications,  ranging from spintronics to cancer therapy in medical science.

\end{abstract}

\date{\today}

\maketitle

\section{Introduction}
Recently, there has been significant interest in two-dimensional (2D) materials because of their fascinating properties and potential application possibilities. Many such materials are currently known. Among them, graphene and other materials like hexagonal boron nitride (h-BN), Black Phosphorus (BP), silicon carbide, and various metal dichalcogenides are the most representative ones~\cite{Geim2013_Grigorieva,lin2010soluble,jafari2022first, weng2016functionalized, na2014few,jafari2020electronic, dahbi2016black, chen2017rising, rahmani2020effect, chabi2021creation, rahmani2021rectifying, komsa2012two, chimene2015two,rahmani2019quantum, wang2018electrical, guo2020electronic,jafari2022electronic}. Currently, the Transition Metal Dichalcogenides (TMDs) are of exceptional interest. These materials are described by the general chemical formula MX$_{2}$, where M stands for a transition metal element (such as Mo, W, Nb, and V), while X represents a chalcogen element (like S, Se, and Te). More than 60 different combinations of these materials are currently known, which have different physical properties: metallic, semiconducting, or superconducting ones~\cite{marseglia1983transition, wilson1969transition, chhowalla2013chemistry, coleman2011two, jafari2022spin}. It is also worth mentioning that the 2D layered TMDs (monolayers, bilayers, or a few-layer structure) usually have properties, especially band structure, that are distinct from those of their bulk counterparts~\cite{jariwala2014emerging}. Many different experimental methods, like exfoliation, liquid exfoliation, and chemical vapor deposition (CVD), have been utilized to achieve these materials in the monolayer or bilayer forms~\cite{wang2012electronics, radisavljevic2011single, lee2010frictional}. 

Most of the transition metal dichalcogenides (TMDs) lack magnetism in their pristine forms, unlike the transition metal trihalides such as CrI3 structures \cite{Xue_applphysrev2021,stagraczynski2024magnetic}.
However, there are methods that enable making them magnetic, like, for instance, doping with magnetic atoms, adsorption of non-metal elements, and leveraging edge effects \cite{shidpour2010density, he2010magnetic, wang2011ultra, li2008mos2, ma2011graphene, ma2011electronic}. For example, the creation of triple vacancies in single-layer MoS$_{2}$ has been proposed as one of the means to introduce a net magnetic moment, while other defects related to Mo and S atoms have a negligible impact on the non-magnetic ground state \cite{xiao2011electrochemically}. 
Generally, controlling the non-magnetic and magnetic states in MoS$_{2}$ nanoribbons can be achieved via the interplay between defects and also adsorption, with the latter depending on the type of defects and adsorption sites. However, doing this in other TMD materials may not be so easy, and one can face more challenges, as exciting magnetic properties in non-magnetic materials are strongly related to the nature of defects, edge states, and dopant positions. 

An intriguing group of the TMDs family is the group of vanadium dichalcogenides (VX$_{2}$, where X can be Sulfur, Selenium, or Tellurium), with VS$_{2}$ being the most prominent representative. These materials exist in two different phases, denoted as H and T, each possessing unique crystal structures and physical characteristics \cite{kan2015density}. Intrinsic ferromagnetism in VS$_{2}$  was firstly predicted using first-principle calculations and then was confirmed experimentally. This achievement received great attention from researchers and put VS$_{2}$ in the range of crucial materials for the development of novel technologies \cite{ma2012evidence, gao2013ferromagnetism}. Recently, one can observe growing interest concerning different stability and magnetic properties of the  H and T phases of VS$_{2}$  monolayers. In contrast to earlier conjectures, the H-phase of these materials is more stable, displaying ferromagnetic order with an easy axis aligned within the layer plane \cite{zhuang2016stability, yue2017first}. 
One should note that the emergence of new experimental techniques that enable the fabrication of VS$_{2}$   monolayers opens the opportunity to develop new nanoelectronics devices  \cite{feng2011metallic, coleman2011two}. There are several experimental techniques that allow to control and modulate the electronic and magnetic properties of VS$_{2}$. One of such methods is the doping strategy \cite{das2015beyond, luo2017structural, ma2012evidence}. Using first-principle calculations, the impact of dopants, like chromium, molybdenum, and tungsten, on the electronic structure and phase transitions of VS$_{2}$ monolayers has been explicated \cite{yadav2023growth}. Furthermore, experimental observations showed that the charge doping could lead to the phase transitions from 2H-phase over 1T phase and subsequently to the emergence of a bipolar magnetic semiconducting behavior, with potential applications in phase-change electrical devices \cite{luo2017structural}. In addition to that, the effect of dopants like Iron (Fe) on strain engineering, as well as magnetic ordering, has been studied to figure out the potential applications in the spintronic field and also enhance the magnetic coupling \cite{ma2012evidence}. The Fe-doping is used to tune the optical, magnetic, and electrical properties of VS$_{2}$, which makes it useful in different fields from electronic to biomedicine \cite{lei2020biodegradable, xiu2022fe, dhakal2022cobalt}. The presence of different Fe-doping structures and their influence on the electronic and magnetic properties of VS$_{2}$ allows to make progress in a wide range of fields, including spintronics, optoelectronics, and catalysis \cite{maibam2022doped}. 
%

In this work, we study VS$_{2}$ and the doping strategy using the computation analysis. We comprehensively analyze the essential properties of the H-phase, and we begin with calculating the electronic properties of the pristine monolayer of VS$_{2}$. Next, we replace V-atoms with the Fe ones in different sites of the extended supercell of VS$_{2}$. The paper is organized as follows: Section \ref{methods} provides an overview of the methodology and computational techniques used. Section \ref{results} discusses the findings derived from the computational analysis. Finally, Section \ref{summary} summarizes the concluding remarks.

\section{Methods}
\label{methods}
To calculate the electronic structure and investigate physical parameters,  we have used the Density Functional Theory (DFT) and the Quantum ATK code package (version 2021.06-SP2) \cite{smidstrup2017first}. This code package relies on the Hohenberg-Kohn theorem \cite{hohenberg1964inhomogeneous}, and Kohn–Sham equations \cite{kohn1965self}. The wave function was expanded using the SG15 collection of optimized norm-conserving Vanderbilt (ONCV) pseudopotentials along with the High Linear Combination of Atomic Orbitals (LCAO-U) basis set. \cite{hamann2013optimized} For the exchange correlation, the Perdew–Burke–Ernzerhof (PBE) generalized-gradient approximation (GGA) was applied. \cite{perdew1996generalized} We set the energy mesh-cutoff to 600 eV and the convergence criteria to 1e-06 eV (1e-08 eV for magnetic anisotropy energy) for each primitive cell. To sample the two-dimensional Brillouin zone, we utilized a k-point grid of 30 × 30 × 1. \cite{monkhorst1976special} All structures under consideration have been fully optimized until the force on each atom dropped below 0.005 eV/Å. To prevent artificial interaction between image layers, we incorporated vacuum layers of at least 20 Angstrom. Furthermore, to account for the electron-electron correlation effect of Vanadium (V) localized 3d orbitals, we implemented DFT+U calculations. Here, U denotes the effective potential of the onsite Coulomb interaction for the Fe-3d or V-3d electrons. \cite{dudarev1998electron}

\begin{figure}
    \centering
\includegraphics[width=1\columnwidth]{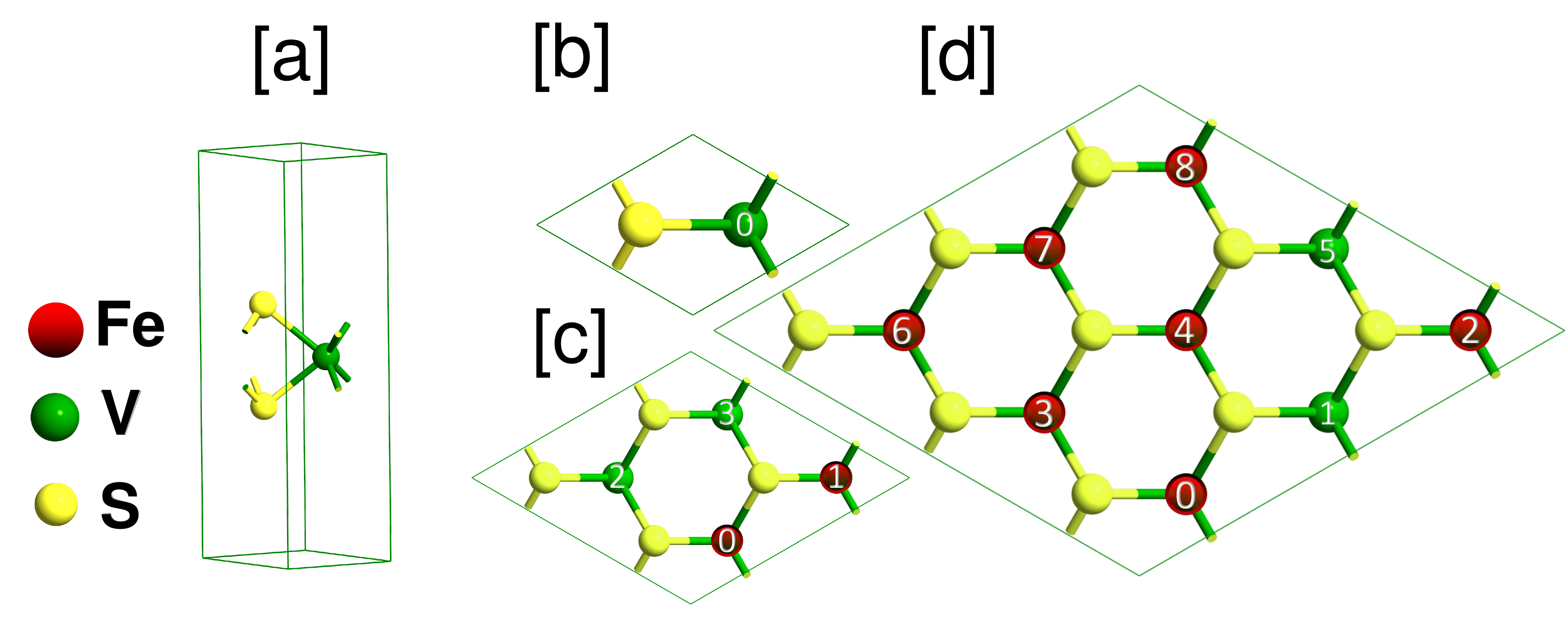}
\caption{Schematic structure of the VS$_{2}$ monolayer without and with Fe-doping: (a) Side view and (b) top view of the pure VS$_{2}$ primitive cell, (c,d) Fe-doping at various sites within 2x2 and 3x3 VS$_{2}$ supercells.}
\label{structures}
\end{figure}

\definecolor{Alto}{rgb}{0.878,0.878,0.878}
\begin{table}
\centering
  \caption{\ Optimized lattice constants $a$, and the bond lengths $d_{V-S}$, $d_{S-S}$, and $d_{Fe-S}$ for various Fe-doped $VS_{2}$ monolayers.}
  \label{table1}
\resizebox{\linewidth}{!}{%
\begin{tblr}{
  cells = {c},
  row{even} = {Alto},
  hlines,
  vlines,
}
\textbf{Types} & \textbf{a[Å]} & $d_{V-S}$[Å] & $d_{S-S}$[Å] & $d_{Fe-S}$[Å]\\
Pure $VS_{2}$ Nanosheet & 3.178 & 2.369 & 2.989 & -\\
Fe-$VS_{2}$ @2x2 @site0 & 6.252 & 2.343 & 2.915 & 2.29\\
Fe-$VS_{2}$ @2x2 @ site1 & 6.250 & 2.342 & 3.043 & 2.29\\
Fe-$VS_{2}$ @3x3 @site0 & 9.462 & 2.357 & 2.930 & 2.30\\
Fe-$VS_{2}$ @3x3 @site3 & 9.462 & 2.353 & 2.986 & 2.30\\
Fe-$VS_{2}$ @3x3 @site4 & 9.461 & 2.353 & 2.986 & 2.30\\
Fe-$VS_{2}$ @3x3 @sites 0,7 & 9.374 & 2.350 & 2.913 & S0=2.28 \textbar{} S7=2.29\\
Fe-$VS_{2}$ @3x3 @sites 2,6 & 9.374 & 2.353 & 2.912 & S2= 2.28 \textbar{} S6= 2.29\\
Fe-$VS_{2}$ @3x3 @sites 3,4 & 9.400 & 2.343 & 2.966 & S3= 2.29 \textbar{} S4= 2.29\\
Fe-$VS_{2}$ @3x3 @sites2,6,8 & 9.330 & 2.370 & 2.853 & S2,8= 2.29 \textbar{} S6= 2.28
\end{tblr}
}
\end{table}

\section{Results} \label{results}
\subsection{Structural Properties}
Various types of Transition Metal Dichalcogenides (TMDs)  exist naturally or have been synthesized, with their structures determined by the way in which the monolayers MX$_{2}$ are arranged. In this formula, M stands for transition-metal elements, like Mo, W, and V, while  X represents chalcogen elements, such as S, Se, and Te. The monolayer of Vanadium diselenide, VS$_{2}$, consists of a hexagonal sheet of V atoms sandwiched between two sheets of chalcogen S atoms. The atomic structure of the V sheets is reminiscent of the hexagonal structure of graphene. 
The VS$_{2}$ monolayer exists in two different forms, i.e., in the  H-phase and in the T-phase. These two phases differ by the way the chalcogen atoms are coordinated. In this paper, we focus on the H-phase as it is more stable. In this phase, with the primitive cells illustrated in Figure \ref{structures}(a) and (b), the nearest six S atoms enclose each V atom, while each S atom is connected to the three nearest-neighbour V  atoms. The V atoms in H-VS$_{2}$ occupy positions within the trigonal prism forms created by the surrounding S atoms. The H-VS$_{2}$ corresponds to the P 6M2 (D3h) symmetry point group, where a V atom sits at Wyckoff site 1(a) with the coordinates (0, 0, 0) and S atoms fill the 2(h) sites at the positions (1/3, 2/3). 
The optimized lattice parameter of pure VS$_{2}$ was found to be 3.178 Å, in good agreement with previous studies \cite{jafari2023electronic,jing2013metallic,feng2011metallic,ma2012evidence}.

Using the optimized VS$_{2}$ monolayer, as described above, we now begin the Fe-doping procedure for assumed supercells and different doping sites. This approach not only allows one to have a comprehensive study of the influence of random Fe-doping, but it is also suitable for comparison with experimental observations, which is crucial for validating theoretical predictions and understanding practical aspects of the materials. 
Figure \ref{structures} (c) illustrates a representative supercell configuration of size 2x2, with Fe-doping at sites labeled as 0 or at sites labeled as 1. The Fe contribution constitutes then 8.33$\%$ of the entire structure. The lattice parameter and bond lengths ($d_{V-S}$, $d_{Fe-S}$) are summarized in Table \ref{table1}. As one can see in this table,  the lattice 
parameter is then reduced by 1.73$\%$ for doping sites labeled as 0.  
This reduction is due to a change in the local atomic environment by the doped Fe atoms, which modifies the pristine crystal lattice and leads to local distortions. The reason for this change is the difference in the atomic radii of Fe and V atoms.
The drop in the lattice constant appears since the iron atom has a smaller atomic radius than the V-atom.  Similarly, other parameters, such as the relevant bond lengths, also exhibit a decrease. Intriguingly, changing the position of Fe from site 0 to site 1 does not induce significant changes in the structural parameters, 
suggesting that the structural integrity of the VS$_{2}$ monolayer is robust against such single Fe-doping perturbations. 

In the next step, we increase the size of the supercell up to the $3\times 3$  primitive unit cells. Then, we introduce Fe atoms in various positions and for different configurations, including single, dual, and triad doping. The contribution of Fe to the entire structure amounts to $3.70\%$, $7.40\%$, and $11.11\%$ for single, dual, and triad Fe-doping, respectively. As in the case of 2x2 supercell,  the lattice parameter becomes reduced with Fe-doping, and for single doping, this reduction is the lowest one and achieves $0.94\%$. The largest reduction, $2.30\%$, is for the triad doping (see Table \ref{table1} for details). Interestingly, dual doping displays a slightly different behavior. When the Fe atoms are positioned far from each other (sites 2 and 6), there is a higher reduction if compared to that when Fe atoms are closer to each other (sites 3 and 4), with the corresponding reductions of $1.88\%$ and $1.57\%$, respectively (see Table \ref{table1}). To account for this difference, we note that 
when the doped Fe atoms are closer, their mutual interaction is stronger, while the overall interaction with the surroundings is reduced to a smaller number of the host-material atoms. This results in a smaller reduction in the lattice parameter of VS$_2$ in comparison to what one would expect if the doped Fe atoms are well separated and independently interact with the neighboring atoms of the host material.
In the following subsections, B and C, we analyze how the structural changes due to Fe doping modify electronic, magnetic, and optical properties.


\begin{figure}[t]
    \centering
\includegraphics[width=1\columnwidth]{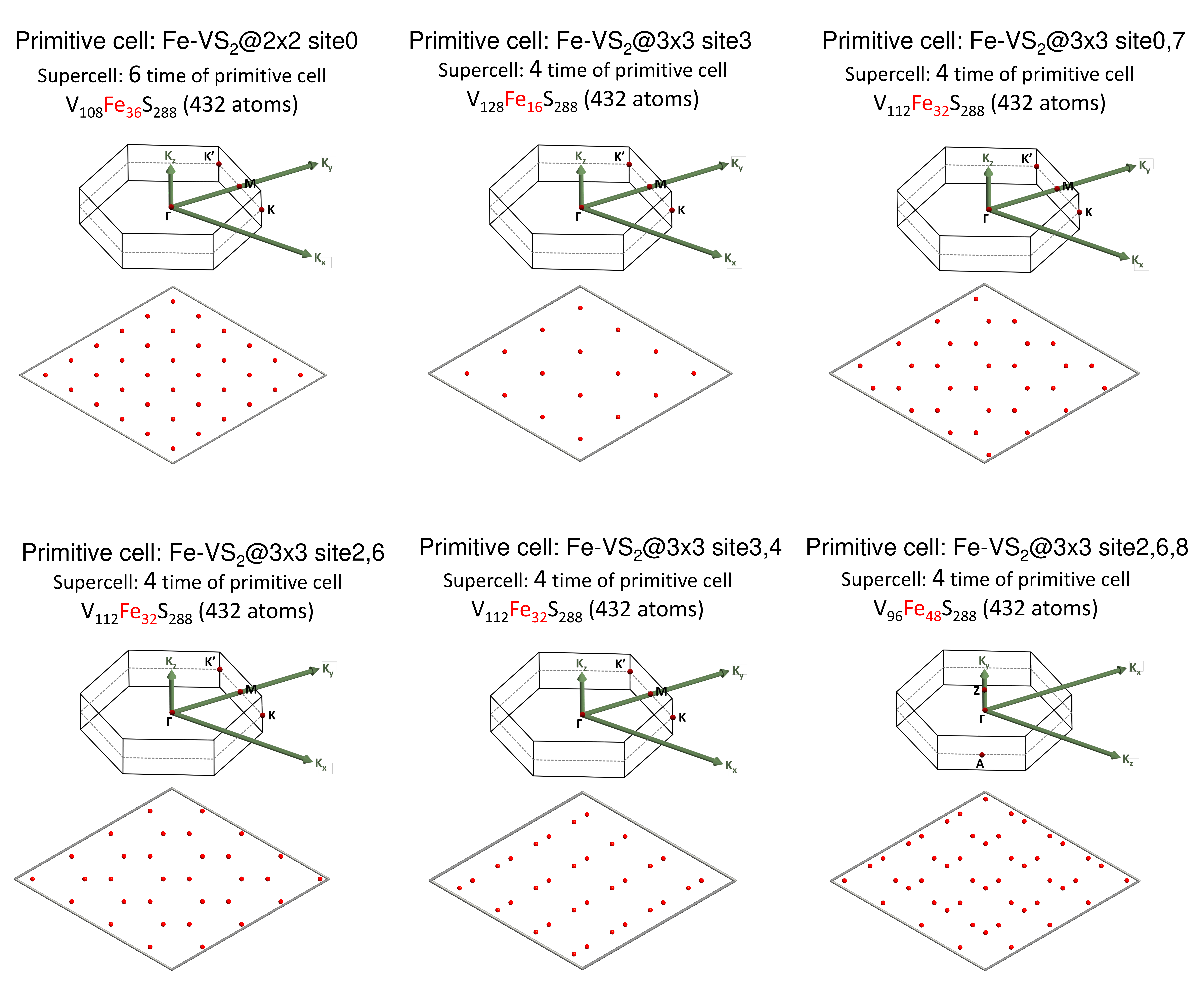}
\caption{Distribution of iron (Fe) atoms according to the supercell and the substituting positions and corresponding Brillouin zones.} 
\label{BZ}
\end{figure}

\subsection{Electronic and Magnetic Properties}
\subsubsection{Electronic bandstructure}

\begin{figure}
    \centering
\includegraphics[width=1\columnwidth]{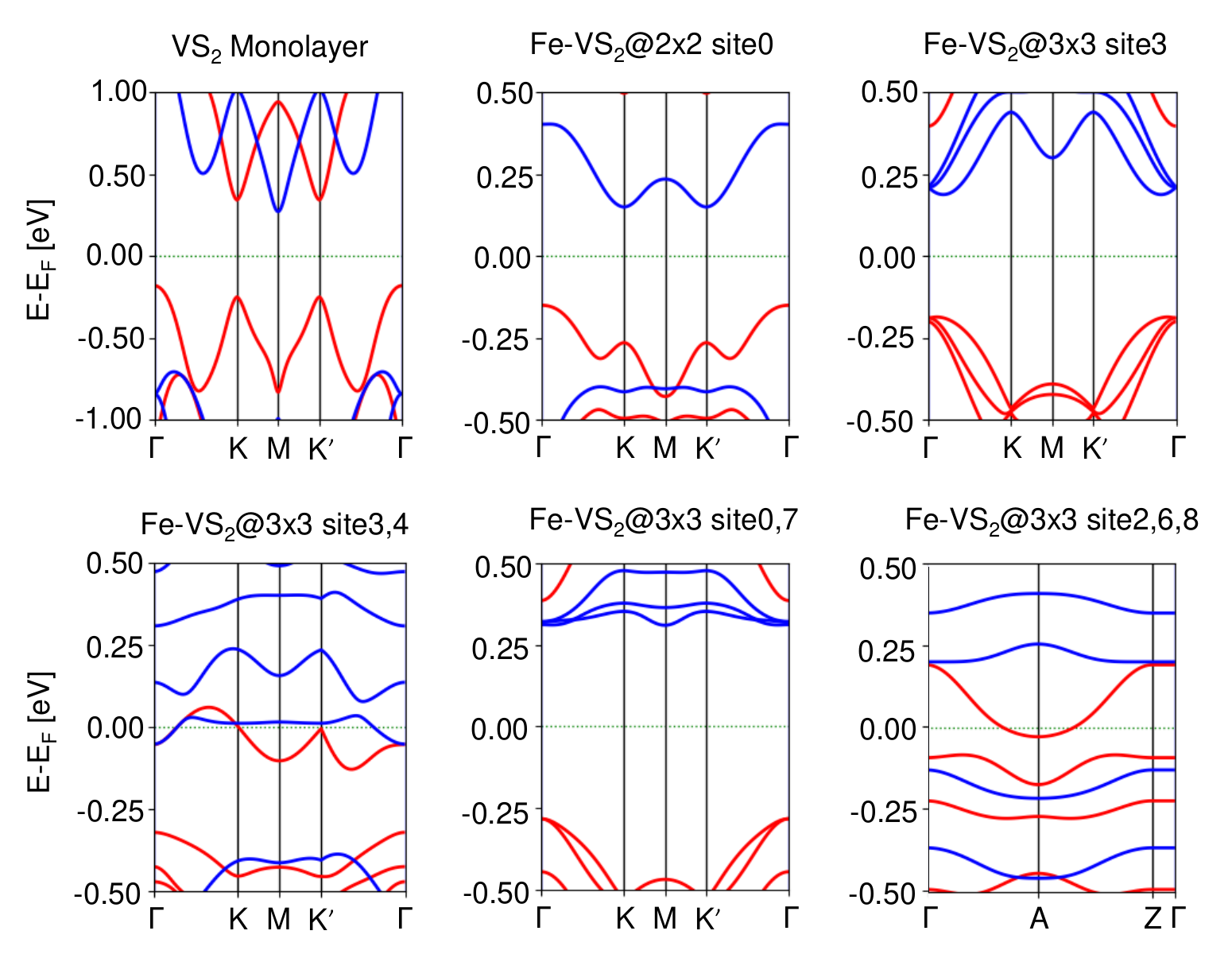}
\caption{Spin-resolved band structures for the pure and Fe-doped VS$_{2}$ monolayer in the PBE+$U$ approximation. The red and blue lines are related to the spin-up and spin-down.}
\label{BS-U}
\end{figure}

\begin{figure*}
    \centering
\includegraphics[width=2 \columnwidth]{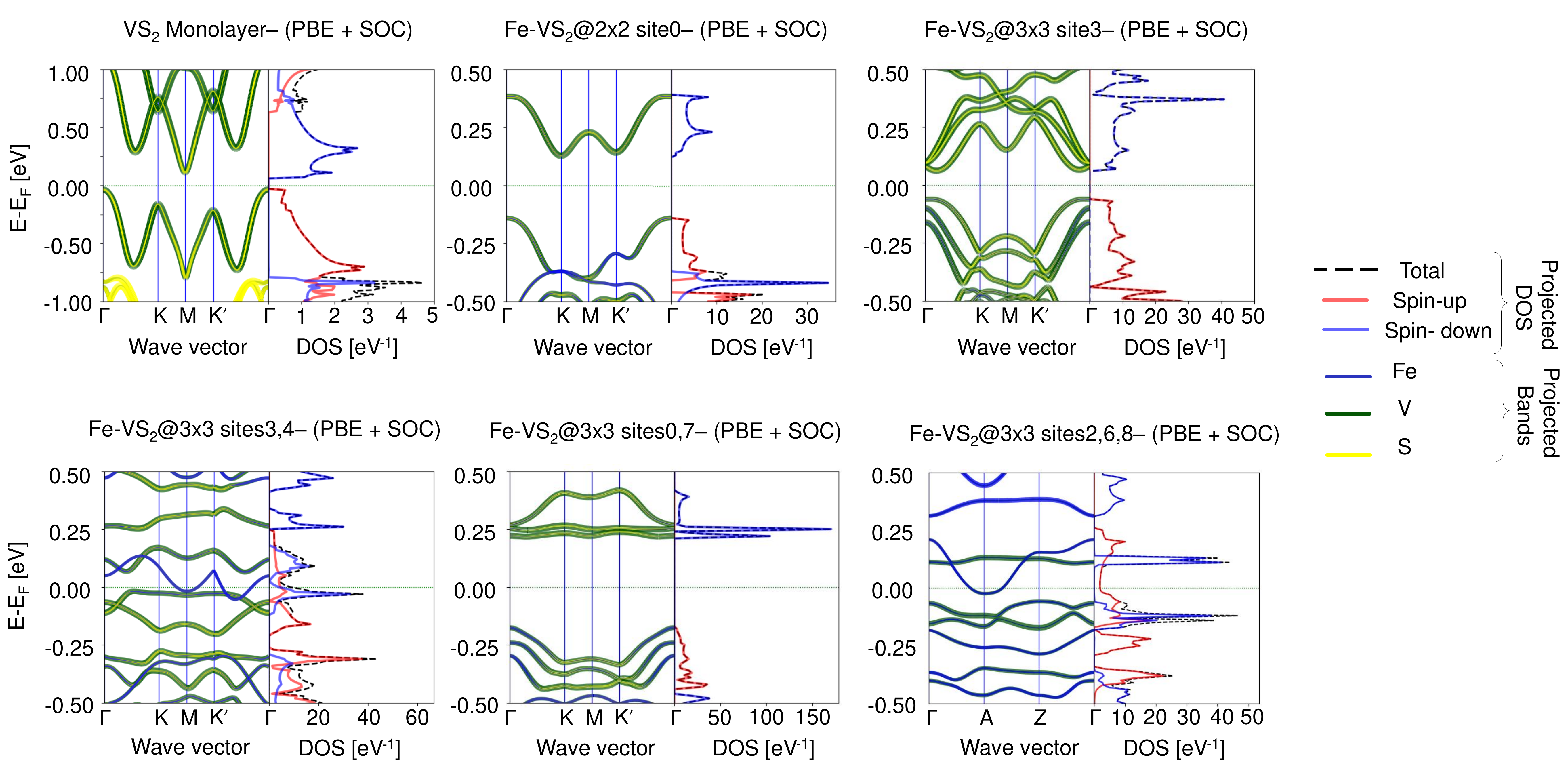}
\caption{Elements-projected band structures and spin projected density of states (DOS) for the pristine and Fe-doped VS$_{2}$ monolayer in the PBE+SOC approximation.}
\label{BS-SOC}
\end{figure*}

To obtain the spin-resolved electronic band structure, we need first to determine the ground state. This is accomplished by examining the total energy difference between various magnetic configurations, including non-magnetic, ferromagnetic (FM), and anti-ferromagnetic (AFM) states. The pure monolayer of VS$_{2}$ exhibits ferromagnetic properties, which is consistent with earlier results~\cite{wu2023ferromagnetic}. It is interesting to note that even with different numbers and random doping positions of Fe atoms, the structures preserve their FM order. This continuity of ferromagnetic behavior with doping appears because iron and vanadium are both transition metals, with their d-orbitals playing the key role in magnetic interactions and stabilization of the ferromagnetic configuration. To get a clearer insight into this issue and make the results more reliable,  we include Coulomb repulsion parameters ($U$) of 1 and 2 eV for Fe and V atoms, respectively~\cite{fuh2016newtype,jafari2024effect}. 
Furthermore, in both the clean VS$_{2}$ monolayer and the Fe-doped structures, there is a spin-orbit coupling (SOC), which is crucial in our analysis. To understand the role of both Coulomb interactions and SOC,  we separately studied their influence on the electronic properties. For band structure calculations,  we selected the $\Gamma$-$K$-$M$-$K'$-$\Gamma$ high symmetry paths within the Brillouin zone (BZ) for all structures.
In the PBE+U calculations, illustrated in Figure \ref{BS-U} and summarized in Table \ref{table-2}, the pure VS$_{2}$ monolayer exhibits a small band gap of 0.45 eV, revealing its semiconducting behavior. Upon including SOC in the calculations, as illustrated in Figure \ref{BS-SOC}, the pure $VS_2$ monolayer keeps its semiconducting behavior, although with slight modifications of the electronic structure observed in the vicinity of the $K$ and $K^{'}$ valleys. These modifications may have some implications for potential applications in electronics, valleytronics, and optoelectronics. 
As the K and K$^{'}$ points are non-equivalent in the presence of SOC, unlike in the absence of spin-orbit coupling, we observe intrinsic valley splitting of 36 meV as illustrated in Figure \ref{BS-SOC} ($VS_2$ monolayer). This valley splitting, $\Delta E$, is notable and also crucial for various electronic properties. Moreover, its magnitude can be modified with different methods such as strain engineering, fabrication methods of heterostructures, and doping strategies. In this study, our objective is to investigate the structural behavior under different doping scenarios across various cell configurations. 

Now, we introduce Fe atoms into the VS$_{2}$ monolayer in various sites and for various supercells. Consider first the case of a 2x2 supercell with the Fe atom doped in site 0, see Figure~\ref{structures}. 
As shown in Figure \ref{BS-U}, the energy bandgap decreases to 0.30 eV upon doping, and this is accompanied by a down-shift of the bands, as it is visible in the points $K$ and $K'$. This effect can be further examined in Figure \ref{BS-SOC}, where we present the projected band structure (FAT-bands) and Projected Density of States (PDOS). Despite the reduction of the bandgap in the PBE+U calculations, the band gap slightly increases when the spin-orbit coupling is included in the calculations. This rather unexpected trend arises from significant modifications in the conduction band induced by the presence of Fe atoms in the 2x2 supercell. Specifically, the localized effect of Fe doping in this configuration likely induces substantial perturbations in the electronic structure of neighboring VS$_{2}$ units. Consequently, this localized impurity has a significant impact on the electronic states, leading to unexpected changes in the band structure under the influence of spin-orbit coupling. Furthermore, the spatial arrangement of Fe atoms in the 2x2 supercell (which is small) leads to a specific electronic interaction that cannot be present in the case of larger supercells.
Thus, the structural and electronic characteristics of the 2x2 supercell, especially concerning the arrangement and interaction of Fe atoms, affect the bandgap when we consider both +U and +SOC calculations. Also noteworthy is the fact that the majority of bands near the Fermi level are associated with VS$_{2}$. However, for valence bands near the points K and K', the influence of Fe doping becomes particularly evident at the K point due to spin-orbit-induced hybridization with the Fe-3d electrons. The hybridization effect can lead to a remarkable difference in the spin polarization and peaks in the DOS near the Fermi level, which comes from the complex interaction between Fe and VS$_{2}$ electronic states.  Additionally, PDOS analysis reveals that the majority of states for the conduction bands are spin-down ones. In contrast, the valence bands predominantly spin up near the Fermi level due to the unique electronic configuration induced by the Fe doping. A similar situation is observed for the Fe doping at site 1 in the 2x2 supercell. Some difference in the band structure appears due to a small difference in the Fe-VS$_{2}$ interactions in the two different doping sites.

\definecolor{Gallery}{rgb}{0.937,0.937,0.937}
\begin{table}
\centering
\caption{Comparison of 
energy bandgap for different approaches of Coulomb and spin-orbit interactions.}
\resizebox{\linewidth}{!}{%
\begin{tblr}{
  cells = {c},
  row{3} = {Gallery},
  row{5} = {Gallery},
  row{7} = {Gallery},
  row{9} = {Gallery},
  row{11} = {Gallery},
  cell{1}{3} = {c=2}{},
  hlines,
  vlines,
}
\textbf{Approach} & \textbf{PBE+U} & \textbf{PBE + SOC} & \\
\textbf{Parameters} & $E_{gap} [eV]$ & $E_{gap} [eV]$ & $MAE [ meV]$\\
Pure $VS_{2}$ Nanosheet & 0.450 & 0.140 & -0.200\\
Fe-$VS_{2}$ @2x2 @site0 & 0.300 & 0.260 & -0.020\\
Fe-$VS_{2}$ @2x2 @ site1 & 0.290 & 0.280 & -0.019\\
Fe-$VS_{2}$ @3x3 @site0 & 0.370 & 0.121 & -1.381\\
Fe-$VS_{2}$ @3x3 @site3 & 0.370 & 0.122 & -1.381\\
Fe-$VS_{2}$ @3x3 @site4 & 0.370 & 0.122 & -1.382\\
Fe-$VS_{2}$ @3x3 @sites 0,7 & 0.590 & 0.392 & -0.806\\
Fe-$VS_{2}$ @3x3 @sites 2,6 & 0.590 & 0.394 & -0.807\\
Fe-$VS_{2}$ @3x3 @sites 3,4 & Metal & Metal & -0.803\\
Fe-$VS_{2}$ @3x3 @sites2,6,8 & Half-Metal & Half-Metal & -0.116
\end{tblr}
}
\label{table-2}
\end{table}

From Table \ref{table-2}, it is evident that increasing the supercell to 3x3 and keeping the single Fe-doping leads to a decrease in the energy band gap, regardless of whether Coulomb or spin-orbit interactions are considered. However, for the case of dual Fe-doping, especially in cases where the Fe atoms are not in nearest neighbor positions (sites 0,7 and sites 2,6), the bandgap increases. Conversely, substituting Fe atoms in two neighboring V sites (sites 3,4) leads to significant geometric modifications compared to the primitive cell structure, resulting in a modified band structure that changes from semiconducting to metallic. A similar trend is observed for triad Fe-doping at sites 2,6,8, where the energy bandgap decreases and the system transforms into a half-metallic state, both in the presence of Coulomb interactions and when the spin-orbit coupling is included. These results are consistent with certain experimental findings. More specifically, experimental  characterization of the band gaps of VS$_{2}$ and Fe-VS$_{2}$ via the solid-state UV spectroscopy has shown that a smaller band gap exists in structures with  
Fe-doping.~\cite{lei2020biodegradable} In our calculations, when a sufficient number of Fe atoms are present compared to the entire VS$_{2}$ monolayer, the bandgap explicitly decreases. This decrease is logical since Fe dopants exist to change the electronic structure and decrease the band gap. However, for the 3x3 supercell at sites 0,7 and sites 2,6, where Fe atoms are not close to each other, the bandgap increases. This behavior is reasonable,  as the reduced interaction between Fe atoms reduces the modifications in the electronic structure and thus increases the bandgap. 

\definecolor{Alto}{rgb}{0.878,0.878,0.878}
\begin{table*}
\centering
  \caption{\ Spin Magnetic Moment (MM) of pure and Fe-doped $VS_{2}$ monolayer for different PBE+U and PBE+SOC approaches. Due to the various numbers of V atoms in different supercells, the highest and lowest MM have been distinguished.}
\resizebox{\linewidth}{!}{%
\begin{tblr}{
  cells = {c},
  row{4} = {Alto},
  row{6} = {Alto},
  row{8} = {Alto},
  row{10} = {Alto},
  row{12} = {Alto},
  cell{1}{2} = {c=5}{},
  cell{1}{7} = {c=5}{},
  cell{2}{1} = {r=2}{},
  cell{2}{2} = {c=5}{},
  cell{2}{7} = {c=5}{},
  cell{3}{4} = {c=3}{},
  cell{3}{9} = {c=3}{},
  cell{4}{2} = {c=2}{},
  cell{4}{4} = {c=3}{},
  cell{4}{7} = {c=2}{},
  cell{4}{9} = {c=3}{},
  cell{5}{2} = {c=2}{},
  cell{5}{4} = {c=3}{},
  cell{5}{9} = {c=3}{},
  cell{6}{2} = {c=2}{},
  cell{6}{4} = {c=3}{},
  cell{6}{9} = {c=3}{},
  cell{7}{4} = {c=3}{},
  cell{7}{9} = {c=3}{},
  cell{8}{4} = {c=3}{},
  cell{8}{9} = {c=3}{},
  cell{9}{4} = {c=3}{},
  cell{9}{9} = {c=3}{},
  cell{10}{5} = {c=2}{},
  cell{10}{10} = {c=2}{},
  cell{11}{5} = {c=2}{},
  cell{11}{10} = {c=2}{},
  cell{12}{5} = {c=2}{},
  cell{12}{10} = {c=2}{},
  vlines,
  hline{1-2,5-14} = {-}{},
  hline{3-4} = {2-11}{},
}
\textbf{Approach} & \textbf{PBE+U} &  &  &  &  & \textbf{PBE+SOC} &  &  &  & \\
\textbf{Parameters } & \textbf{Magnetic Moment $[\mu_{B}]$} &  &  &  &  & \textbf{Magnetic Moment $[\mu_{B}]$} &  &  &  & \\
 & \textbf{Max. of V} & \textbf{Min. of V} & \textbf{Fe} &  &  & \textbf{Max. of V} & \textbf{Min. of V} & \textbf{Fe } &  & \\
Pure $VS_{2}$ Nanosheet & 1.26 &  & - &  &  & 1.17 &  & - &  & \\
Fe-$VS_{2}$ @2x2 @site0 & All sites = 0.57 &  & 0.52 &  &  & site2 = 0.51 & site3 = 0.48 & site0 = 0.30 &  & \\
Fe-$VS_{2}$ @2x2 @ site1 & All sites = 0.57 &  & 0.52 &  &  & site2 = 0.58 & site0 = 0.58 & site1 = 0.52 &  & \\
Fe-$VS_{2}$ @3x3 @site0 & site7 = 1.30 & site1 = 1.03 & 1.09 &  &  & site7 = 1.20 & site1 = 0.92 & site 0 = 0.81 &  & \\
Fe-$VS_{2}$ @3x3 @site3 & site1 = 1.30 & site7 = 1.03 & 1.09 &  &  & site1 = 1.20 & site7 = 0.92 & site3 = 0.81 &  & \\
Fe-$VS_{2}$ @3x3 @site4 & site2 = 1.30 & site5 = 1.03 & 1.09 &  &  & site2 = 1.20 & site5 = 0.92 & site4 = 0.81 &  & \\
Fe-$VS_{2}$ @3x3 @sites 0,7 & site5 = 1.27 & site3 = 0.68 & site0 = 0.78 & site7 = 0.65 &  & site5 = 1.17 & site3 = 0.53 & site0 = 0.43 & site7 =0.35 & \\
Fe-$VS_{2}$ @3x3 @sites 2,6 & site4 = 1.27 & site7 = 0.68 & site2 = 0.78 & site 6 = 0.65 &  & site4 = 1.16 & site7 = 0.53 & site2 = 0.43 & site 6 = 0.34 & \\
Fe-$VS_{2}$ @3x3 @sites 3,4 & site1 = 0.93 & ~site7 = 0.23 & site3 = 0.37 & site4 = 0.37 &  & site1 = 0.67 & site7 = 0.10 & site3 =0.38 & site4 =0.33 & \\
Fe-$VS_{2}$ @3x3 @sites2,6,8 & site0 = 0.53 & site7 = 0.16 & site2 = 0.39 & site 6 = 0.84 & site8 = 0.86 & site4 = 0.44 & site3 = 0.07 & site2 = 0.44 & site 6 = 0.76 & site8 = 0.75
\end{tblr}
}
\label{table-3}
\end{table*}

\begin{figure}
    \centering
\includegraphics[width=1\columnwidth]{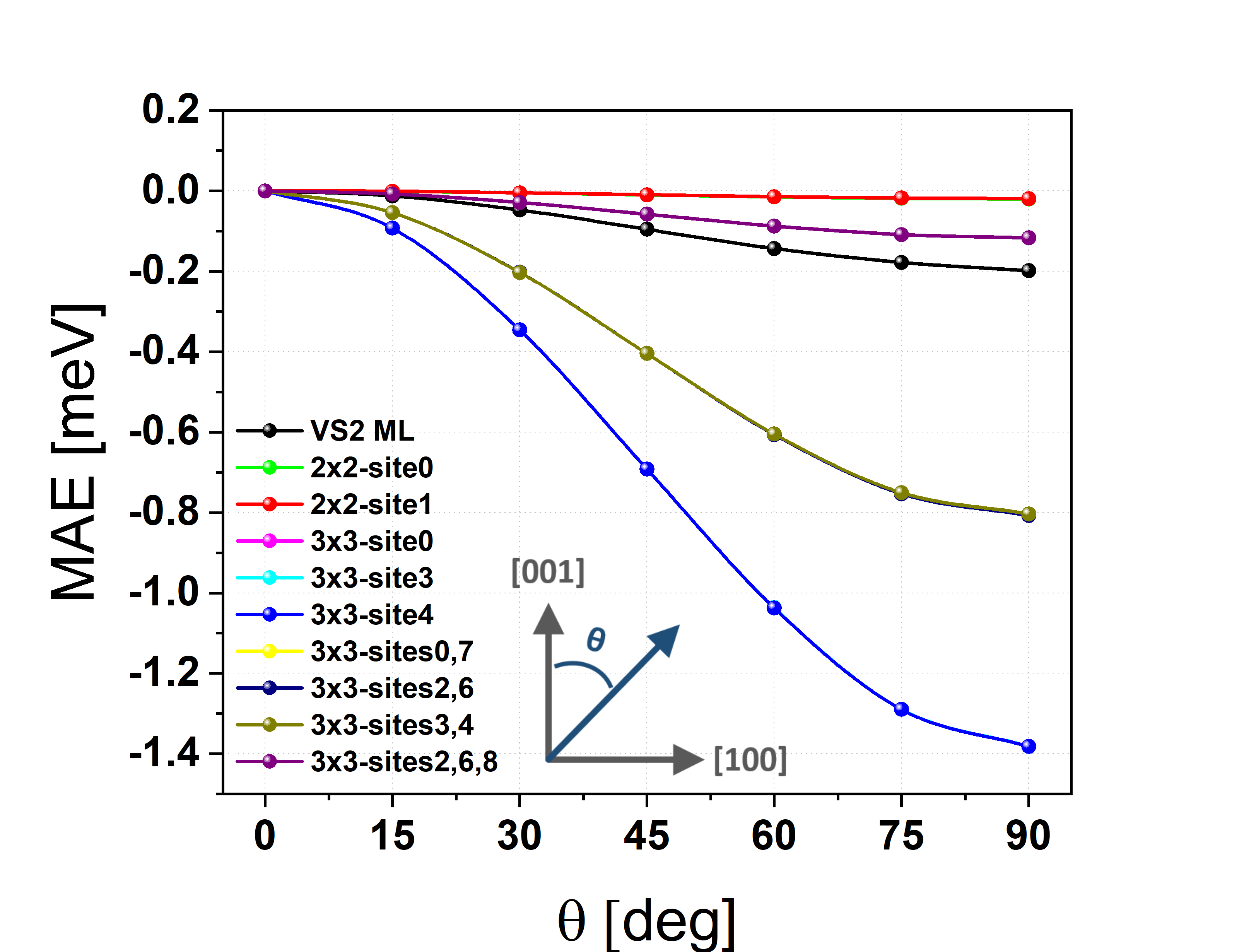}
\caption{Angular dependence of Magnetic Anisotropy Energy (MAE) per unit cell of perfect $VS_{2}$ monolayer (ML) with and without Fe-doping and for indicated supercell sizes and doping concentrations. The MAE is presented as a function of the polar angle $\theta$, with the azimuthal angle set to 0 (see the inset for the definition of $\theta$).}
\label{MAE}
\end{figure}

\subsubsection{Magnetic Anisotropy Energy and Magnetic Moment}
One of the important parameters required to understand the behavior of magnetic materials is the Magnetic Anisotropy Energy (MAE). Such an anisotropy determines the preferred orientation of magnetization in the ground state and plays a crucial role in different applications of magnetic systems, e.g., in spin valves, magnetic data storage, magnetic sensors, and others. The magnetic anisotropy arises from the spin-orbit interaction, which makes the energy of an electron dependent on the orientation of its magnetic moment. This, in turn, results in the emergence of preferred magnetic orientations within the system.
Magnetic anisotropy can be determined by calculating the energy difference between two relevant spin orientations using the force theorem, which gives insight into the stability and behavior of magnetic systems. 
Generally, the energy difference between two well-defined magnetic states, i.e., MAE,  can be calculated from the formula  \cite{daalderop1990first}:
\begin{equation}
\resizebox{0.9\columnwidth}{!}{$
MAE = \sum _i f_i(\theta _1, \phi _1) \epsilon _i(\theta _1, \phi _1) - \sum _i f_i(\theta _0, \phi _0) \epsilon _i(\theta _0, \phi _0), 
$}
\end{equation}
where, $f_i(\theta, \phi)$ denotes the occupation factor for band $i$ (including the $k$-point index), with $(\theta, \phi)$ representing the spin orientation and $\epsilon_i(\theta, \phi)$ indicating the corresponding band energy. The indices 0 and 1 at $(\theta$ and $\phi)$ refer to the two corresponding magnetic orientations (two magnetic states). In this paper, we focus on the perpendicular (out-of-plane or easy plane) and in-plane anisotropies. In the case where the system is magnetically isotropic (or nearly isotropic) within the plane, the perpendicular anisotropy is defined as $\text{MAE}=E_{[100]}-E_{[001]}$ (or $\text{MAE}=E_{[010]}-E_{[001]}$), with the $x$ and $y$ axes lying within the layer plane and the $z$-axis perpendicular to the layer. Therefore, a positive value of MAE indicates a perpendicular easy axis, whereas a negative value corresponds to a perpendicular hard axis, i.e., to easy-plane anisotropy. As it follows from Table \ref{table-2} and also as illustrated in Figure \ref{MAE}, the pure monolayer of $VS_{2}$ exhibits an easy-plane anisotropy of MAE $\approx$ 0.20 meV.
However, when we increase the supercell size to 2x2 with single Fe doping, it is clearly seen that the easy plane anisotropy survives, but its magnitude is remarkably reduced, and this reduction is directly related to the presence of Fe-doping.
It is interesting to note that in the case of supercell increased to 3x3, the  magnitude of the easy-plane anisotropy increases 
to -1.381 meV for single-site doping and to -0.80 meV for dual doping. 
In turn, in the presence of triad Fe doping, the easy plane magnetic anisotropy is reduced again to  -0.11 meV, but it is larger than that for the 2x2 supercell with single-site doping. 
We note, that the ability to control the magnetic anisotropy energy is interesting not only from the fundamental side, but also from practical reasons.


In our calculations on the Fe-doped VS$_{2}$ monolayers, we applied two different approaches to study their magnetic properties -- including PBE+U and PBE+SOC ones. The results on magnetic moments are listed in Table \ref{table-3}. Firstly, we investigated the magnetic moment of V atoms in the pristine VS$_{2}$ monolayer, where we found 1.26 $\mu_{B}$ and 1.17 $\mu_{B}$  using the PBE+U and PBE+SOC methods, respectively. Then, by increasing the supercell size to a 2x2 one and doping Fe atoms, a sharp decrease is observed in the magnetic moment of V atoms. This is due to significant changes in the local electronic structure and magnetic interaction within the VS$_{2}$ lattice, which originate from the presence of Fe atoms.  In turn, the magnetic moment of the Fe atoms in $VS_2$ is reduced as compared to its free-standing value, which indicates the interactions with VS$_{2}$. In turn,  in the case of the 3x3 supercell, the magnetic moments are remarkably different for different Fe doping sites. For certain sites, we find an enhanced magnetic moment, while in other sites, it is reduced. 
This variety is directly related to the interaction between doped Fe-atoms and the VS$_{2}$ lattice.
These findings emphasize the complexity of magnetic phenomena in Fe-doped $VS_{2}$ monolayers and provide valuable insights for the potential applications.

\subsection{Optical Properties}
An important objective of this work is the analysis of tuning optical properties of  VS$_{2}$   by doping it with Fe atoms. 
To achieve this objective, we employ the Kubo-Greenwood formula to calculate the susceptibility tensor \cite{sipe1993nonlinear,harrison1970solid}:
\begin{equation}
\chi_{ij}(\omega) = \frac{e^2}{\hbar m_e^2 V} \sum_{nm\mathbf{k}} \frac{(f_{m\mathbf{k}}-f_{n\mathbf{k}}) p_{nm}^i(\mathbf{k}) p_{mn}^j(\mathbf{k})}{\omega_{nm}^2(\mathbf{k}) [\omega_{nm}(\mathbf{k})- \omega - i \Gamma/\hbar]},
\end{equation}
where $p_{nm}^i$ denotes the matrix element of the $i$-th component of the momentum operator between the states $n$ and $m$, $m_e$ represents the electron mass, $e$ is the electron charge, $V$ stands for volume, $\Gamma$ is the energy broadening, $\hbar\omega_{nm} = E_n - E_m$, and $f_{n\mathbf{k}}$ indicates the Fermi distribution function evaluated at the band energy $E_n(k)$.

The response coefficients, i.e., the relative dielectric function $\epsilon (\omega )$, polarizability $\alpha  (\omega )$, and optical conductivity $\sigma  (\omega )$ are connected to the susceptibility as follows \cite{probert2011electronic}:
\begin{align}
\epsilon(\omega) &= (1 + \chi(\omega) ), \\
\alpha(\omega) &= V \epsilon_0 \chi(\omega), \\
\sigma(\omega) &= - i \omega \epsilon_0 \chi(\omega),
\end{align}
where $\epsilon_0$ is the dielectric constant.
The dielectric function is a complex quantity,  $\epsilon(\omega) = \epsilon_{r}(\omega) + i \epsilon_{i}(\omega)$, where the first and second terms are the real and imaginary parts, respectively. 
%
In turn, the refractive index $n$ is linked with the complex dielectric constant through the formula:
\begin{equation}
n + i \kappa = \sqrt{\epsilon},
\end{equation}
where $\kappa$ denotes the extinction coefficient. In terms of the real ($\epsilon_r$) and complex  ($\epsilon_i$) parts of the dielectric function, the refractive index $n$ and extinction coefficient $\kappa$ are expressed as \cite{martin2020electronic}:
\begin{align}
n &= \sqrt{\frac{\sqrt{\epsilon_r^2+\epsilon_i^2}+\epsilon_r}{2}}, \\
\kappa &= \sqrt{ \frac{ \sqrt{ \epsilon_r^2+ \epsilon_i^2} - \epsilon_r}{2}}.
\end{align}
The optical absorption coefficient $\alpha_a$ is associated with the extinction coefficient as \cite{griffiths2023introduction}:

\begin{equation}
\alpha_a = 2 \frac{\omega}{c} \kappa ,
\end{equation}
while
the reflectivity $R$ is given by \cite{desjarlais2005density}:
\begin{equation}
R = \frac{(1-n)^2 + \kappa^2}{(1+n)^2 + \kappa^2},
\end{equation}

\begin{figure}
    \centering
\includegraphics[width=1\columnwidth]{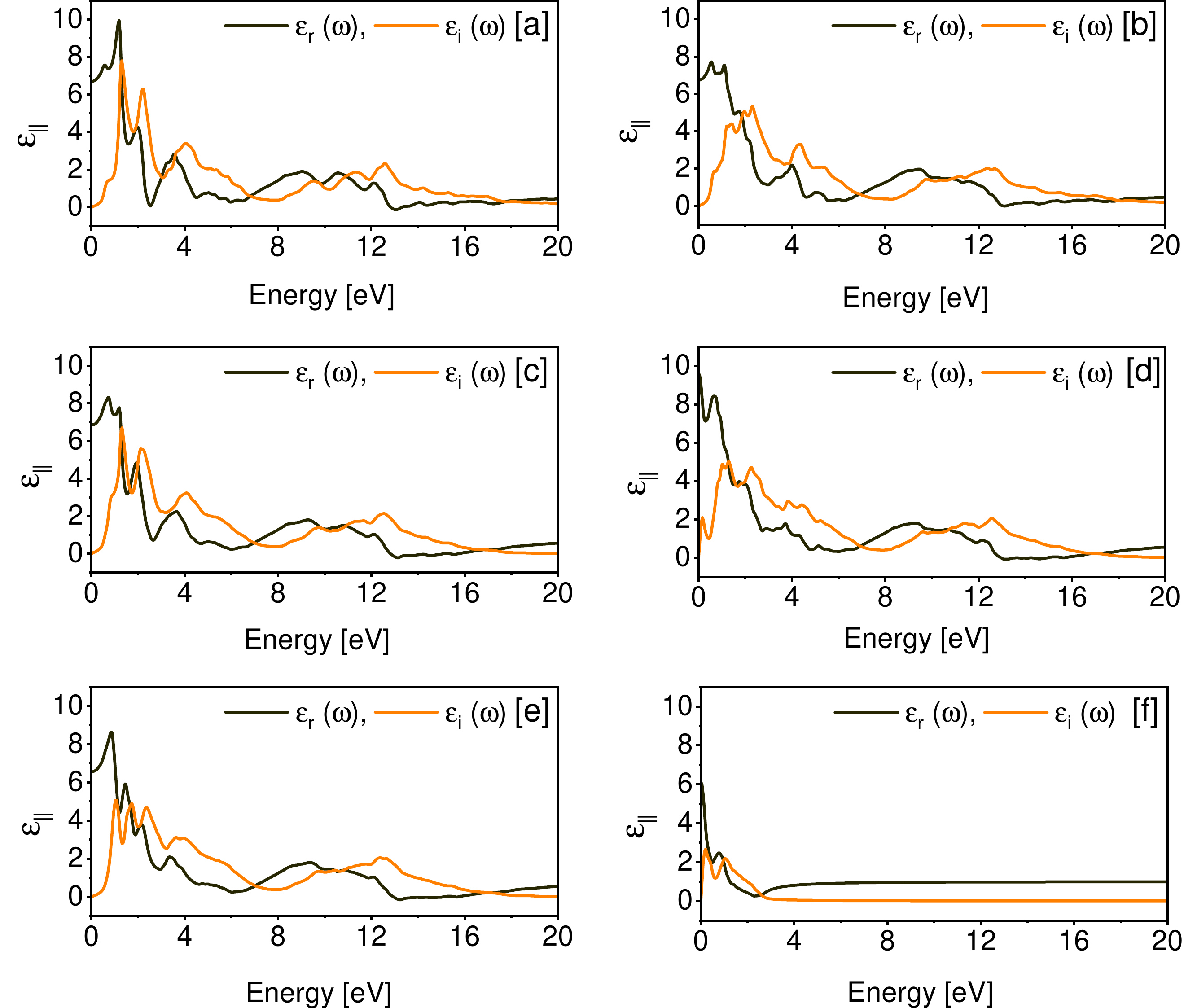}
\caption{Real part, $\epsilon_{r}$, and imaginary part, $\epsilon_{i}$, of the dielectric function for selected structures as a function of energy for: (a) pure VS$_2$ monolayer, (b) Fe-VS$_2$ @2x2@site0, (c) Fe-VS$_2$ @3x3@site3, (d) Fe-VS$_2$ @3x3@sites3,4, (e) Fe-VS$_2$@3x3@sites0,7, and (f)  
Fe-S$_2$@3x3@sites2,6,8.} 
\label{dielectric}
\end{figure}

The calculated real and imaginary parts of the dielectric function of $VS_2$ for selected doped structures are shown in Figure \ref{dielectric}. 
One can note distinct trends that show fundamental electronic transitions within the materials. Both the real, $\epsilon_r$ and imaginary, $\epsilon_i$, parts exhibit prominent peaks in the energy range around 1 eV, see Figure \ref{dielectric}(a), which suggest significant electronic excitations within this energy range. The sharp drop following the peaks and subsequent fluctuations at higher energies reflect the complex interactions between the electronic states and band structure effects inherent to these materials. When the Fe atoms are not strongly interacting with each other,  in most cases, one can see trends similar to those in the pristine $VS_2$ monolayer. However, the introduction of Fe atoms into the $VS_2$ lattice in the geometries (3x3 sites-3,4) and (3x3 sites-2,6,8) leads to pronounced changes in the dielectric response, particularly observed in the real part of the dielectric constant. In contrast to the gradual increase observed in the case of pure $VS_2$ monolayer, the dielectric constant of the Fe-doped $VS_2$ monolayer starts with a peak in $\epsilon_r$, Figure \ref{dielectric}d and Figure \ref{dielectric}f. This is due to the modifications in the electronic structure arising from the presence of the Fe atoms, which lead to changes in the band gaps and/or hybridization effects between Fe-d orbitals and the $VS_2$ host lattice states, where in the band structures, it is clearly seen that the band gap disappears and the structure changes to metallic or half-metallic. Such modifications not only influence the material's optical properties but also have the potential for applications in optoelectronic devices.
\begin{figure}
    \centering
\includegraphics[width=0.6\columnwidth]{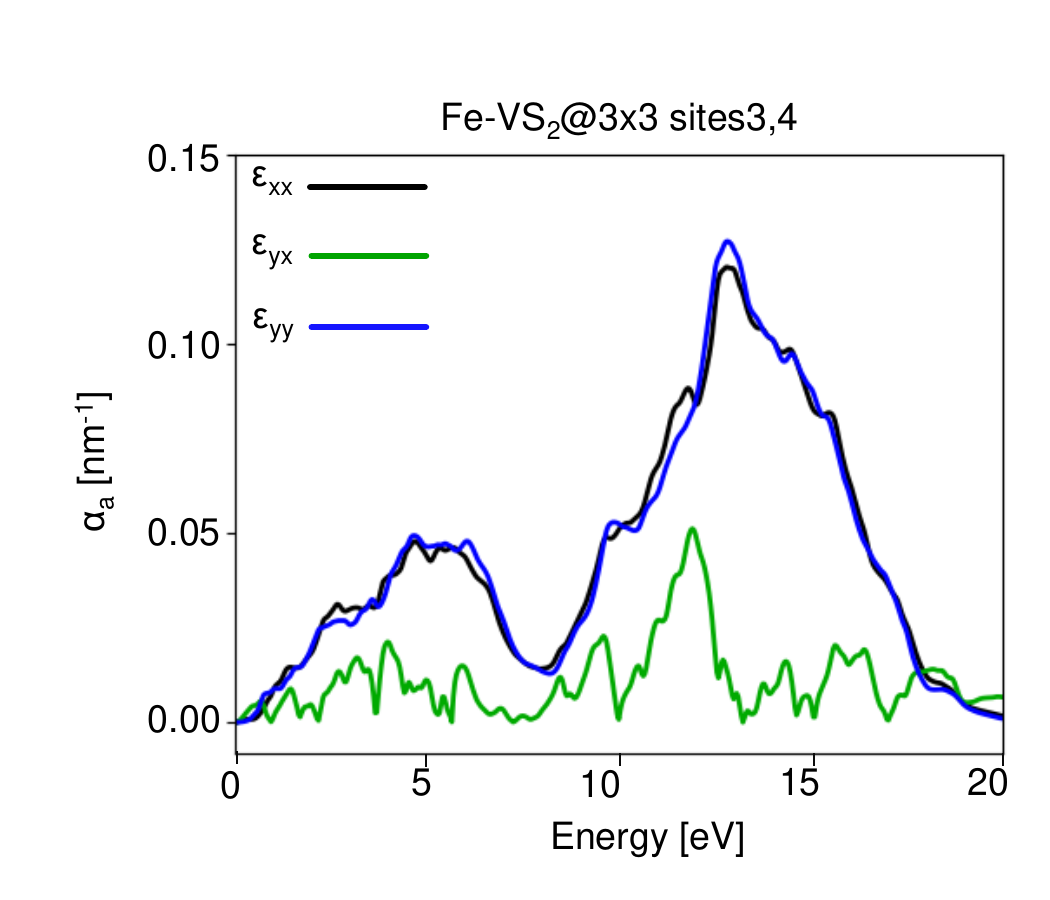}
\caption{Absorption coefficient of Fe-doped $VS_{2}$ in specific case of supercell 3x3@sites 3,4 and in different crystal's directions of $\epsilon_{xx}$, $\epsilon_{yx}$, and $\epsilon_{yy}$.}
\label{absorption}
\end{figure}

Further analysis of the absorption coefficient, $\alpha$, for the structure Fe-VS$_2$ (3x3 sites-3,4) reveals the anisotropic optical properties.
Unlike the isotropic response observed in the pure $VS_2$ monolayer and in most other Fe-doped structures,  in this particular case, the Fe-doped monolayer exhibits different values of $\epsilon_{xx}$, $\epsilon_{yy}$, and $\epsilon_{yx}$, see Figure \ref{absorption} for details. This anisotropy reflects the preferential alignment or orientation of the Fe dopants within the atomic lattice.
A comprehensive study of the dielectric function and absorption coefficient can provide valuable insight into the electronic and optical properties of both the pristine and Fe-doped $VS_2$.  The results achieved in this paper allow for  deeper understanding of the fundamental properties of the studied $VS_2$ monolayers, and also can facilitate finding the ways for  their practical  applications.
\section{Summary and conclusions}
\label{summary}
In this paper, we have studied by the DFT methods both pristine and Iron-doped $VS_2$ monolayers. We focused on the modifications of electronic and magnetic properties by various doping schemes. To determine the electronic structure, we assumed that the doped systems are periodic with the relevant supercell. Numerically, we have considered the cases when doping was in the primitive cell of $VS_2$, as well as for the supercells that are 2x2 and 3x3 times larger than the primitive cell. We have considered situations with different distributions of the doped Fe atoms in the supercells. Both electron correlation parameter $U$ and spin-orbit coupling were included in the calculations.  

Having determined the electronic structure, we have analyzed the magnetic moments of the V and Fe atoms and the dependence of these moments on the distribution of impurities within the supercells. In addition, using the Kubo Greenwood formula, we have calculated optical properties, including dielectric function, reflectivity, and extinction coefficient of the investigated structures. The obtained results show that the optical properties can be modulated by selective doping of  $VS_2$ monolayers. They also show that the doped structures have properties that are of potential applications in nanoelectronics, spintronics, and optoelectronics.

\section*{Author contributions}
\textbf{Mirali Jafari:} Conceptualization, Methodology, Software, Validation, Formal analysis, Investigation, Data Curation, Writing - Original Draft, Visualization \textbf{Nasim Rahmani-Ivriq:}Conceptualization, Formal analysis, Writing - Original Draft \textbf{Anna Dyrdał:} supervision, writing – review \& editing.

\section*{Conflicts of interest}
There are no conflicts to declare.

\section*{Data availability}
Data for this article, including computational results used for band structure calculations (PBE+U and PBE+SOC), magnetic anisotropy energy (MAE), dielectric constant ($\epsilon_{\parallel}$), and absorption spectra ($\alpha_a$), are available on Zenodo at \href{https://doi.org/10.5281/zenodo.14165791}{https://doi.org/10.5281/zenodo.14165791}.



\section*{Acknowledgements}
This work has been supported by the Norwegian Financial Mechanism 2014 - 2021 under the {Polish -- Norwegian} Research Project NCN GRIEG “2Dtronics” no. 2019/34/H/ST3/00515.

\bibliography{ref.bib}

\end{document}